\documentclass[twocolumn,superscriptaddress,amssymb,10pt,pra,floatfix]{revtex4}

\usepackage{graphicx}
\usepackage{dcolumn}
\usepackage{bm}
\usepackage{epsfig}
\usepackage{color}
\usepackage{longtable}
\date{December 24, 2016}

\usepackage[cp866]{inputenc}
 \usepackage[english]{babel}
 \newcommand{\be}{\begin{equation}}
\newcommand{\ee}{\end{equation}}

\newcommand{\bea}{\begin{eqnarray}}
\newcommand{\eea}{\end{eqnarray}}

\begin{document}

\preprint{}
%
%
%
%
\title{Relativistic vortex electrons: \\ paraxial versus non-paraxial regimes}
%
%
%
%

\author{Dmitry Karlovets}
\affiliation{Tomsk State University, Lenina Ave. 36, 634050 Tomsk, Russia}

\date{\today \\[0.3cm]}

%
%
%
%
%
\begin{abstract}
A plane-wave approximation in particle physics implies that a width of a massive wave packet $\sigma_{\perp}$ is much larger than its Compton wavelength $\lambda_c = \hbar/mc$.
For Gaussian packets or for those with the non-singular phases (say, the Airy beams), corrections to this approximation are attenuated as $\lambda_c^2/\sigma_{\perp}^2 \ll 1$
and usually negligible. Here we show that this situation drastically changes for particles with the phase vortices associated with an orbital angular momentum $\ell\hbar$.
For highly twisted beams with $|\ell| \gg 1$, the non-paraxial corrections get $|\ell|$ times enhanced and $|\ell|$ can already be as large as $10^3$.
We describe the relativistic wave packets, both for vortex bosons and fermions, which transform correctly under the Lorentz boosts, are localized in a 3D space, 
and represent a non-paraxial generalization of the massive Laguerre-Gaussian beams. We compare such states with their paraxial counterpart paying specific attention to relativistic effects
and to the differences from the twisted photons. In particular, a Gouy phase is found to be Lorentz invariant and it generally depends on time rather than on a distance $z$. 
By calculating the electron packet's mean invariant mass, magnetic moment, etc., we demonstrate that the non-paraxial corrections can already reach the relative values of $10^{-3}$. 
These states and the non-paraxial effects can be relevant for the proper description of the spin-orbit phenomena in relativistic vortex beams, of scattering of the focused packets by atomic targets, of collision processes in particle and nuclear physics, and so forth.
\end{abstract}
%
%
%
%

\maketitle

\section{Introduction}

Many properties of the massive particles carrying orbital angular momentum (OAM) can be described within the models of a non-localized Bessel beam (akin to a plane wave) \cite{Bliokh, Review} 
and of a localized Laguerre-Gaussian packet \cite{Review}. While the latter is applicable only in the paraxial regime, the former also predicts a number of non-paraxial effects for relativistic vortex electrons, 
such as a spin-orbit interaction, thanks to a finite transverse momentum. However, the Bessel beam is unsuitable for the problems in which finite sizes of the wave packet and its finite energy-momentum dispersions become important.

Depending on a problem, this may happen when the beam is focused to a spot of a size $\sigma_{\perp}$ comparable to a Bohr radius, $a \approx 0.053$ nm, or to an electron's Compton wavelength, $\lambda_c = \hbar/mc \approx 0.39$ pm. As the vortex electrons have already been focused to a spot of $\sigma_{\perp} \gtrsim 0.1$ nm $\approx 2a$ \cite{Angstrom},
a more realistic wave-packet approach beyond the paraxial approximation is needed, in particular, for proper study of the spin-orbit phenomena
and for scattering problems in atomic and high-energy physics, especially when the quantum interference and coherence play a notable role \cite{Sarkadi, Sch, PRL}.
The quantum interference of the incoming packets is of crucial importance -- say, for potential applications in hadronic physics \cite{JHEP} -- and the orthogonal Bessel states do not describe these effects. 

In collisions of particles described as wave packets instead of the simplified plane waves, non-paraxial corrections to the plane-wave cross-sections 
are generally attenuated as $\lambda_c^2/\sigma_{\perp}^2 \ll 1$ and do not play any essential role. 
Nevertheless, these corrections can be enhanced when the OAM of a projectile is large, $|\ell|\hbar \gg \hbar$ \cite{JHEP}. 
As a result, for well-focused highly twisted beams (the current record is $|\ell| \sim 10^3$ \cite{l1000}) these effects can compete with the two-loop corrections to the QED processes like $e^{-}\gamma \rightarrow e^{-}\gamma, e^-e^- \rightarrow e^-e^-$, etc. 

Precise quantitative estimates of these phenomena require that the vortex wave packets be spatially localized, described in a Lorentz invariant way, 
and applicable beyond the paraxial approximation. Despite the recent interest in the relativistic wave packets with OAM \cite{Barnett, Birula}, 
such a model is still lacking (see the recent discussion in \cite{Bliokh17,Review}) first and foremost because the widely used coordinate representation is much less convenient for these purposes 
than the momentum one. The situation here is somewhat reminiscent of that with a photon wave function in x-space \cite{Birula_photon}.

Furthermore, for massive particles in a $(3+1)$-D space-time the very definition of the paraxial regime needs to be revised, as the characteristic scale, the Compton wavelength $\lambda_c$, 
is Lorentz invariant, unlike a de Broglie wavelength, $\lambda_{dB} = 2\pi/p$, or the wavelength of a massless photon. As we argue in this paper, an invariant condition of paraxiality, $\sigma_{\perp} \gg \lambda_c$ (or $\sigma \ll m$ in momentum space), is more appropriate for relativistic description of the vortex electrons than the commonly used non-invariant one, $p_{\perp}\ll p_z$.

The massive wave packets are actively used in neutrino physics \cite{Akhmedov_09, Akhmedov_10} and their Lorentz invariant description 
has been recently given by Naumov and Naumov in \cite{Packets1, Packets2, Naumov}.
In a widespread model, the packet is Gaussian in $p$-space, it has a mean 4-momentum $\bar{p}_{\mu}$ (a parameter), $\bar{p}^2 = m^2$, and an uncertainty $\sigma$, 
which is a Lorentz scalar and vanishing, $\sigma \ll m$, in the paraxial regime. Here we further develop this approach  by adding OAM to the set of quantum numbers, both for bosons and fermions,
and treating the non-paraxial effects. Such vortex packets are localized in a 3D space and, what is crucial for relativistic phenomena, transform correctly under the Lorentz boosts. 
In the paraxial approximation, these packets reduce to the invariant Laguerre-Gaussian beams, which have notable differences from their photonic counterpart due to the massiveness of the packet
(for relativistic energies they -- of course -- vanish).

We calculate such observables as the vortex electron's mean energy, magnetic moment, etc., and demonstrate that the non-paraxial effects are $|\ell|$ times \textit{enhanced} 
for such vortex packets compared to the Bessel beam, to the Gaussian packet or even to that with a non-singular phase (say, an Airy packet \cite{Airy, Airy_El_Exp}). 
For instance, a corresponding correction to an invariant mass of the electron packet is positive and with current technology it can reach the values of $(10^{-4}-10^{-3})\,m$. 
To put it differently, such a wave packet is $0.01\%-0.1\%$ heavier than an ordinary plane-wave electron. 
It is this weighting that can reveal itself in the corresponding invariant corrections, 
$$
\sim |\ell| \lambda_c^2/\sigma_{\perp}^2 \gtrsim \alpha_{em}^2 = 1/137^2 \gg \lambda_c^2/\sigma_{\perp}^2,
$$ 
to the $e^-e^-$ or $e^-\gamma$ scattering.

Thus, the sub-nm-sized highly twisted beams can be used for probing the previously unexplored non-paraxial effects in the high-energy and nuclear physics,
analogously to the recently described quantum coherence phenomena in atomic physics \cite{Sarkadi, Sch} and in addition to the magnetic-moment effects in electromagnetic radiation \cite{PRL13}. 
Moreover, in scattering of a coherent superposition $|\ell_1\rangle + |\ell_2\rangle$ of two vortex electrons by atoms, the analogous fundamental scale is the Bohr radius $a$, 
which is $1/\alpha_{em} = 137$ times larger than $\lambda_c$. As a result, the corresponding effects may become only moderately attenuated (akin to \cite{PRL}), 
and one would need an explicitly non-paraxial approach, feasible with the packets described in this paper.

Here we study only the azimuthally symmetric packets with a vanishing mean transverse momentum, that is, $\bar{\bm p} = \{0,0,\bar{p}\}$. 
Hence we imply only the longitudinal boosts; see \cite{Bliokh17trans} for effects of the transverse ones. 
When calculating the wave functions $\psi^{\text{par}} (x)$ in the paraxial regime, the following expansion for the energy is employed:
\begin{eqnarray}
& \displaystyle
\varepsilon = \sqrt{{\bm p}^2 + m^2} \approx \bar{\varepsilon} + \bar{{\bm u}} ({\bm p} - \bar{\bm p}) + \cr
& \displaystyle + \frac{1}{2\bar{\varepsilon}} \left (\delta_{ij} - \bar{u}_i \bar{u}_j \right) ({\bm p} - \bar{\bm p})_i({\bm p} - \bar{\bm p})_j,\cr
& \displaystyle
\bar{{\bm u}} = \bar{\bm p}/\bar{\varepsilon},\ \bar{\varepsilon} = \sqrt{\bar{\bm p}^2 + m^2}.
\label{eps}
\end{eqnarray}
This expansion does not alter the transformational properties of the resultant packets with respect to the Lorentz boosts.
The $4$-vector $\bar{p}^{\mu} = \{\bar{\varepsilon}, 0, 0, \bar{p}\}$ serves as a parameter of the state and it is only in the paraxial regime that the packet's mean $4$-momentum $\langle p_{\mu}\rangle$ 
coincides with $\bar{p}_{\mu}$. A system of units $\hbar = c = e = 1$ is used and the metric is $g_{\mu\nu} = \text{diag}(1,-1,-1,-1)$, so that $\bar{p}_{\mu} x^{\mu} = \bar{\varepsilon} t - z\bar{p}$.

\section{Non-relativistic vortex packet}

We start with a non-relativistic wave packet with OAM, which is a good benchmark example the subsequent relativistic vortex packets can be compared with. 
In a model of a Gaussian beam with a phase $\varphi ({\bm p})$, the mean momentum $\bar{\bm p}$, and with the momentum uncertainty $\sigma$, the wave function is
\begin{eqnarray}
& \displaystyle \psi ({\bm p},t) = \Big ({\frac{2 \sqrt{\pi}}{\sigma}}\Big )^{3/2}\, \exp \Big \{-it\varepsilon -\frac{({\bm p} - {\bm {\bar p}})^2}{2\sigma^2} + i \varphi ({\bm p})\Big\},\cr
& \displaystyle  \int\frac{d^3p}{(2\pi)^3}\, |\psi ({\bm p},t)|^2 = 1,  
\label{psi}
\end{eqnarray}
where $\varepsilon = {\bm p}^2/2m$. When the phase $\varphi ({\bm p})$ represents a smooth and analytical function (for instance, for the Airy beams \cite{Airy, Airy_El_Exp}), 
neither the probability density nor the observables like energy depend on the phase. 
For a vortex beam, however, the phase $\varphi_{\ell} =\ell \phi_p$ is not defined at the point ${\bm p}_{\perp} = 0$ 
and its derivative,
\begin{eqnarray}
\frac{\partial \varphi_{\ell}}{\partial {\bm p}} = \ell\, \frac{\hat{\bm z} \times {\bm p}}{{\bm p}_{\perp}^2},\ \hat{\bm z} = \{0,0,1\},
\label{dvarphi}
\end{eqnarray}
diverges when ${\bm p}_{\perp} \rightarrow 0$. To put it another way, the Fourier transform of $\psi ({\bm p},t)$ decays non-exponentially, namely as $1/\rho^2$, at large distances. 

In order to restore exponential decay of the vortex electron's wave function with $\rho$,
one can add a prefactor of $p_{\perp}^{|\ell|}$ to $\psi({\bm p},t)$ and modify the normalization constant accordingly\footnote{Interestingly, the transverse part of this wave function,
$\psi_{\ell}({\bm p},t) \propto \frac{(p_{\perp}/\sigma)^{|\ell|}}{\sqrt{|\ell|!}}\, \exp\{-p_{\perp}^2/(2\sigma^2)\}$, formally coincides with an overlap between a coherent state $|\alpha\rangle$ 
and a photon number state $|n\rangle$, $\langle n|\alpha\rangle = \frac{\alpha^n}{\sqrt{n!}}\, \exp\{-|\alpha|^2/2\}$.}:
\begin{eqnarray}
& \displaystyle \psi_{\ell}({\bm p},t) = \Big ({\frac{2\sqrt{\pi}}{\sigma}}\Big )^{3/2}\frac{p_{\perp}^{|\ell|}}{\sigma^{|\ell|}\sqrt{|\ell|!}}\cr
& \displaystyle \times\exp \Big \{-it\varepsilon -\frac{({\bm p} - \bar{\bm p})^2}{2\sigma^2} + i\ell\phi_p\Big\},\cr
& \displaystyle \int\frac{d^3p}{(2\pi)^3} |\psi_{\ell}({\bm p},t)|^2 = 1,  
\label{psiell}
\end{eqnarray}
and now $|\psi_{\ell}({\bm p},t)|^2$ does depend on $|\ell|$, as for the Bessel beam.

A Fourier transform of this state can be calculated exactly:
\begin{eqnarray}
& \displaystyle \psi_{\ell}({\bm r}, t) = \int\frac{d^3p}{(2\pi)^3}\, \psi_{\ell}({\bm p}, t)\, e^{i{\bm p}{\bm r}} = \cr
& \displaystyle = \frac{1}{\pi^{3/4}\sqrt{|\ell|!}\,\sigma^{|\ell| + 3/2}}\,\frac{(i\rho)^{|\ell|}}{(\sigma^{-2} + it/m)^{|\ell| + 3/2}}\cr 
& \displaystyle \times \exp \Big \{-i \bar{\varepsilon} t + i \bar{\bm p} z + i\ell\phi_r  -\frac{1}{2}\frac{\rho^2 + (z-\bar{u}t)^2}{\sigma^{-2} + it/m}\Big\},\cr
& \displaystyle \int d^3x\, |\psi_{\ell}({\bm r}, t)|^2 = 1,  
\label{psiellx}
\end{eqnarray}
where $\bar{\varepsilon} = \bar{\bm p}^2/2m,\, \bar{{\bm u}} = \bar{\bm p}/m$,
and which is a simplest Laguerre-Gaussian packet with a radial index $n=0$, as a generalized Laguerre polynomial $L_{n=0}^{|\ell|}(x) = 1$. 

The transverse part of the probability density,
\begin{eqnarray}
& \displaystyle |\psi_{\ell}({\bm \rho}, z=t=0)|^2 = \text{const}\,\frac{(\rho\sigma)^{2|\ell|}}{|\ell|!}\,e^{-(\rho\sigma)^2},  
\label{Poissnonrel}
\end{eqnarray}
represents \textit{a gamma distribution} (or a Poisson distribution for $|\ell|$; see, for instance, \cite{Mandel}) with its typical doughnut-like profile and one maximum (because $n=0$). 

Note that within the non-relativistic framework Eq.(\ref{psiellx}) represents \textit{an exact solution} of the Schr\"odinger equation and, in contrast to the twisted photons, 
its Gouy phase, 
$$
\propto \arctan (t\sigma^2/m),
$$ 
depends on the time $t$ and \textit{not} on the distance $z$. This is ultimately due to the massiveness of the packet and, more importantly, 
this feature also holds in the relativistic regime in which the corresponding states are only approximate (paraxial) solutions to the Klein-Gordon equation (see Sec.\ref{LG}). 

Finally, it is also worth noting that by using the light-cone (null-plane) variables $t \pm z$ the corresponding relativistic states can still satisfy the Klein-Gordon equation \textit{exactly} (see, for instance, \cite{PRA15, Bagrov, Bagrov_Mono}). On the other hand, the latter formalism requires a physical justification and for the Laguerre-Gaussian modes 
it may even suffer from some inconsistencies (discussed, for instance, in \cite{Birula}). That is why we shall not deal with it in this paper, 
especially because the \textit{non-paraxial} wave packets with OAM, which are exact solutions either of the Klein-Gordon equation or of the Dirac one,
can readily be obtained in the conventional variables.

\section{Relativistic boson}

\subsection{Phaseless wave packet}

Let us proceed with a massive boson and normalize its phaseless and Lorentz invariant wave function $\psi (p)$ by the following condition
\begin{eqnarray}
& \displaystyle 
\int \frac{d^3p}{(2\pi)^3}\frac{1}{2\varepsilon}\,|\psi (p)|^2 = 1,
\label{bosonnorm}
\end{eqnarray}
which is invariant too and where $\varepsilon = \sqrt {{\bm p}^2 + m^2}$. Here we follow an approach of Refs.\cite{Packets1, Packets2, Naumov}, albeit with another normalization 
(somewhat simplified models of this kind were previously studied, for instance, in \cite{Kowalski}).
The wave function characterized by a 4-scalar $\sigma$ and by the 4-momentum (parameter) $\bar{p}_{\mu}$ can depend only on a scalar $(p_{\mu}-\bar{p}_{\mu})^2 \equiv (p-\bar{p})^2 \leq 0$ (or upon $-(p_{\mu}+\bar{p}_{\mu})^2$) where 
$$
p^2 = \bar{p}^2 = m^2.
$$
Drawing an analogy to Eq.(\ref{psi}), we define\footnote{The integrals we deal with are $\int_{-\infty}^{\infty}dx\, \{1, \cosh x\} \exp\{-\omega \cosh x\} = 2 \{K_0(\omega), K_1(\omega)\},\, \text{Re} \omega > 0$ \cite{Gr}.}
\begin{eqnarray}
& \displaystyle \psi (p) = \frac{2^{3/2}\pi}{\sigma}\, \frac{e^{-m^2/\sigma^2}}{\sqrt{K_1(2m^2/\sigma^2)}}\,\exp\left\{\frac{(p-\bar{p})^2}{2\sigma^2}\right\},
\label{psiboson}
\end{eqnarray}
which is real and satisfies (\ref{bosonnorm}). Here, $K_1$ is a modified Bessel function.

For vanishing mass, $m \ll \sigma$, this wave function vanishes as $\psi (p) \propto m/\sigma \rightarrow 0$, 
that is why it seems that this formalism cannot be directly applied to the massless particles (say, to photons). On the other hand, it works perfectly in the ultrarelativistic limit, $\bar{\varepsilon} \approx \bar{p} \gg m$, as the corresponding wave functions are Lorentz invariant. This allows one to make a comparison between the nearly massless bosons and photons\footnote{If needed, one can get rid of this vanishing behavior by normalizing the wave function to $\sigma^2/2m^2$ instead of unity. In this case, the function itself stays finite in the limit $m\rightarrow 0$ but its normalization tends to infinity, which demands that the mass be finite, though arbitrarily small, exactly as in (\ref{psiboson}). These subtleties seem purely technical (somewhat analogously to 
a problem of the photon wave function \cite{Birula_photon}), first and foremost because the very one-particle description becomes inapplicable when $\sigma \gg m$ ($\sigma_{\perp} \ll \lambda_c$).}. 

In the paraxial regime with $\sigma \ll m$, one gets 
\begin{eqnarray}
& \displaystyle
K_{\ell} (\chi \gg 1) = e^{-\chi} \sqrt{\frac{\pi}{2}} \Big (\chi^{-1/2} + \frac{(2\ell)^2-1}{8}\,\chi^{-3/2} + \cr
& \displaystyle + \mathcal O (\chi^{-5/2})\Big ),\, \chi \equiv 2m^2/\sigma^2,
\label{Kapp}
\end{eqnarray}
where the first correction depends on $\ell$, and so
\begin{eqnarray}
& \displaystyle \psi (p) \simeq \left (\frac{2\sqrt{\pi}}{\sigma}\right)^{3/2}\sqrt{2m}\,\Big (1 - \frac{3}{32}\frac{\sigma^2}{m^2} + \cr
& \displaystyle + \mathcal O (\sigma^4/m^4)\Big )\exp\left\{\frac{(p-\bar{p})^2}{2\sigma^2}\right\}.
\label{psibosonparax}
\end{eqnarray}

Here, both the uncertainty $\sigma$ and the condition $\sigma \ll m$, are Lorentz invariant, 
which may seem to contradict the model of Ref.\cite{JHEP} in which a symmetric $3\times 3$-matrix $\sigma_{ij}$ is used instead.
In the latter approach, the longitudinal component of this matrix experiences a Lorentz extension, $\sigma_{zz} = \bar{\varepsilon} \sigma_{zz}^{\prime}/m$,
which is quite natural. However, a close scrutiny reveals equivalence of these two approaches in the paraxial regime, that of \cite{Naumov} and that of \cite{JHEP}. 
Indeed, let us expand the energy $\varepsilon$ in (\ref{psibosonparax}) in a vicinity of $\bar{\bm p}$ (see Eq.(\ref{eps})). 
With an accuracy up to the second order terms, we get
\begin{eqnarray}
& \displaystyle \frac{(p-\bar{p})^2}{2\sigma^2} = - \frac{1}{2\sigma^2} \left (\delta_{ij} - \bar{u}_i \bar{u}_j\right) ({\bm p} - \bar{\bm p})_i({\bm p} - \bar{\bm p})_j.
\label{psibosonparax2}
\end{eqnarray}
Compared to the non-relativistic expression (\ref{psi}), this one has a new term $\bar{u}_i \bar{u}_j$, and it is this term that provides correct Lorentz transformation of the exponent. 
For a packet moving along the $z$ axis on average, we get
\begin{eqnarray}
& \displaystyle - \frac{1}{2\sigma^2} \left (\delta_{ij} - \bar{u}_i \bar{u}_j\right) ({\bm p} - \bar{\bm p})_i({\bm p} - \bar{\bm p})_j = \cr
& \displaystyle = - \frac{{\bm p}_{\perp}^2}{2\sigma^2} - \frac{m^2 (p_z - \bar{p})^2}{2\bar{\varepsilon}^2\sigma^2},
\label{psibosonparax3}
\end{eqnarray}
which is invariant under a longitudinal boost, as the segment in the momentum space exhibits Lorentz extension, $(p_z - \bar{p})^2 \sim \bar{\varepsilon}^2/m^2$ (so that $\Delta p \Delta x = \text{inv}$). Comparing this with Eq.(3.45) from \cite{JHEP}, we find a matrix $\sigma_{ij}^{-2} = \sigma^{-2} (\delta_{ij} - \bar{u}_i \bar{u}_j)$, which transforms as prescribed,
and $\text{det}\, \sigma_{ij} = \sigma^3 \bar{\varepsilon}/m$. The latter equality provides invariance of the normalization adopted in \cite{JHEP}, 
although only in the paraxial regime.

A Fourier transform of the non-paraxial wave function (\ref{psiboson}) can be found exactly
\begin{eqnarray}
& \displaystyle \psi (x) = \int \frac{d^3 p}{(2\pi)^3} \frac{1}{2\varepsilon}\, \psi(p)\, e^{-ipx} = \cr
& \displaystyle  = \frac{1}{\pi\sqrt{2}}\,\frac{\sigma}{\varsigma}\, \frac{K_1 \left(\varsigma m^2/\sigma^2\right)}{\sqrt{K_1\left(2m^2/\sigma^2\right)}},\cr
& \displaystyle \varsigma = \frac{1}{m}\sqrt{\left (\bar{p}_{\mu} + ix_{\mu} \sigma^2\right )^2} = \text{inv},\, \text{Re}\,\varsigma > 0.
\label{psibosonx}
\end{eqnarray}
Obviously, it satisfies the Klein-Gordon equation and also vanishes as $\psi (x) \propto m \rightarrow 0$ in the zero-mass limit. We call this function a \textit{non-paraxial} wave function of the bosonic packet, even though it does not obey any simple Lorentz invariant normalization condition, as the zeroth component of the current $j^0$ is not positively defined (see, for instance, \cite{BLP}). 
One simple way to check whether the function (\ref{psibosonx}) is properly normalized 
is to calculate the packet's mean energy by using the energy-momentum tensor $T_{\mu\nu}$:
\begin{eqnarray}
& \displaystyle \langle \varepsilon\rangle = \int d^3x\, T^{00} = \int d^3x \Big (\partial_0 \psi^*(x) \partial_0 \psi(x) + \cr
& \displaystyle + {\bm \nabla}\psi^*(x)\cdot{\bm \nabla}\psi(x) + m^2 |\psi(x)|^2\Big) =\cr
& \displaystyle = \int\frac{d^3p}{(2\pi)^3}\frac{1}{2\varepsilon}\, |\psi(p)|^2\, \varepsilon,
\label{energymean}
\end{eqnarray}
as should be.
The integral can also be evaluated exactly,
\begin{eqnarray}
\displaystyle \langle \varepsilon\rangle = \bar{\varepsilon}\, \frac{K_2\left(2m^2/\sigma^2\right)}{K_1\left(2m^2/\sigma^2\right)} = \bar{\varepsilon}\,\left(1 + \frac{3}{4} \frac{\sigma^2}{m^2} + \mathcal O \left(\frac{\sigma^4}{m^4}\right)\right).
\label{energymeanex}
\end{eqnarray}
Note that it is also because of these subtleties with the normalization in x-space that the momentum representation turns out to be more convenient in the non-paraxial regime than the coordinate one. 

Analogously, we find the mean momentum,
\begin{eqnarray}
& \displaystyle \langle {\bm p}\rangle^i = \int d^3x\, T^{i0} = \int\frac{d^3p}{(2\pi)^3}\frac{1}{2\varepsilon}\, |\psi(p)|^2\, {\bm p}^i =
\cr & \displaystyle = \bar{\bm p}^i\, \frac{K_2\left(2m^2/\sigma^2\right)}{K_1\left(2m^2/\sigma^2\right)}.
\label{momentummeanex}
\end{eqnarray}
As a result, an invariant mass of such a wave packet,
\begin{eqnarray}
& \displaystyle m_{\text{inv}}^2 = \langle \varepsilon\rangle^2 - \langle {\bm p}\rangle^2 = m^2\, \frac{K_2^2\left(2m^2/\sigma^2\right)}{K_1^2\left(2m^2/\sigma^2\right)} = \cr
& \displaystyle = m^2 \,\left(1 + \frac{3}{2} \frac{\sigma^2}{m^2} + \mathcal O (\sigma^4/m^4)\right),
\label{minvwp}
\end{eqnarray}
is bigger than the mass $m$ of the plane-wave electron. 

The non-paraxial correction 
$$
\sigma^2/m^2 \equiv \lambda_c^2/\sigma_{\perp}^2 = \mathcal O(\hbar^2)
$$ 
is less than $10^{-6}$ for modern electron microscopes ($\sigma_{\perp} > 0.1$ nm) and less than $10^{-14}$ for modern electron accelerators ($\sigma_{\perp} > 1\,\mu \text{m}$), 
although it can reach the values of the order of $10^{-11}$ for the next-generation colliders like CLIC and ILC for which one of the beam's transverse sizes will be of the order of $1$ nm \cite{PDG}. 

Unlike the ``dynamic'' corrections due to the quantum recoil and spin in scattering or radiation, which are $\mathcal O(\hbar)$, 
this ``kinematic'' term describes purely quantum corrections $\mathcal O(\hbar^2)$ to the particle motion. As we work with the one-particle positive-energy states 
and the corrections to energy and mass are positive (which also holds for a vortex fermion below), 
these effects have no relation to the so-called Zitterbewegung.

Due to (\ref{Kapp}), the non-paraxial wave function (\ref{psibosonx}) decays exponentially at large spatial distances $\sqrt{-x_{\mu}^2} \rightarrow \infty$:
\begin{eqnarray}
& \displaystyle \psi(x)_{\sqrt{-x_{\mu}^2} \rightarrow \infty} \propto \frac{1}{(-x_{\mu}^2)^{3/4}}\, \exp\left\{-\frac{\sqrt{-x_{\mu}^2}}{\lambda_c}\right\}.
\label{psiass}
\end{eqnarray}
with a characteristic scale (``a packet's width'') being the Compton wavelength. 

\begin{figure}[t]
	\center
	\includegraphics[width=0.9\linewidth]{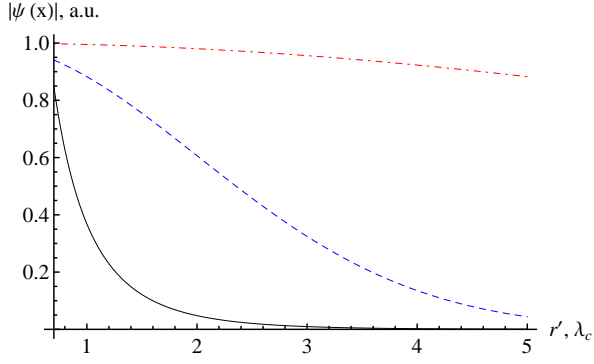}
	\caption{The spatial dependence at large distances and $t=0$ of the exact non-paraxial wave packet (\ref{psibosonx}) (black solid line) and of the paraxial one (\ref{psibosonparaxx}) (blue and red lines). 
The blue dashed line: $\sigma_{\perp} = 2 \lambda_c$, the red dash-dotted line: $\sigma_{\perp} = 10 \lambda_c$. The width of the non-paraxial packet does not exceed a few Compton wavelengths.}
\label{FigR}
\end{figure}

The different behavior takes place in an approximate paraxial formula for $\psi(x)$, 
which is found by expanding the energy in the exponent of Eq.(\ref{psiboson}) according to Eq.(\ref{eps}) and evaluating the corresponding Gaussian integral. 
The result is
\begin{eqnarray}
& \displaystyle \psi^{\text{par}} (x) = \frac{1}{(\sigma\sqrt{\pi})^{3/2} \sqrt{2m}} \frac{1}{(1/\sigma^2 + it/\bar{\varepsilon})^{3/2}}\cr
& \displaystyle \times \exp\Big\{-i\bar{p}_{\mu} x^{\mu} - \frac{1}{2} \frac{1}{1/\sigma^2 + it/\bar{\varepsilon}}\,\Big(\delta_{ij} + \cr
& \displaystyle + \frac{\bar{u}_i \bar{u}_j}{1-\bar{{\bm u}}^2}\Big )({\bm r} - \bar{{\bm u}}t)_i ({\bm r} - \bar{{\bm u}}t)_j\Big\},
\label{psibosonparaxx}
\end{eqnarray}
and it is an approximate solution to the Klein-Gordon equation. This expression is also Lorentz invariant, as 
$$
t/\bar{\varepsilon} = \tau/m = \text{inv}
$$
where $\tau$ is the boson's proper time, and for a longitudinal boost:
\begin{eqnarray}
& \displaystyle
\left(\delta_{ij} + \frac{\bar{u}_i \bar{u}_j}{1-\bar{{\bm u}}^2}\right )({\bm r} - \bar{{\bm u}}t)_i ({\bm r} - \bar{{\bm u}}t)_j = \cr
& \displaystyle = \rho^2 + \bar{\varepsilon}^2 (z-\bar{u}t)^2/m^2 = ({\bm r}^{\prime})^2 = \text{inv}
\label{rprop}
\end{eqnarray}
with ${\bm r}^{\prime}$ being the observation vector in a rest frame of the boson. Unlike the exact state (\ref{psibosonx}), the paraxial one $\psi^{\text{par}} (x)$ is normalized by a simple invariant formula,
\begin{eqnarray}
& \displaystyle \int d^3 x\, 2\bar{\varepsilon}\, |\psi^{\text{par}} (x)|^2 = 1.
\label{paraxnorm}
\end{eqnarray}
Such a packet is much wider than the non-paraxial one (\ref{psibosonx}), as its behavior is Gaussian,
\begin{eqnarray}
& \displaystyle \psi^{\text{par}} ({\bm r}^{\prime},t=0) \propto \exp\left\{-\frac{1}{2}\frac{({\bm r}^{\prime})^2}{\sigma_{\perp}^2}\right\},
\label{psibosonparaxxass}
\end{eqnarray}
where the beam width $\sigma_{\perp} \equiv 1/\sigma \gg \lambda_c$. We compare the exact law of Eq.(\ref{psiass}) with the approximate one of Eq.(\ref{psibosonparaxxass}) in Fig.\ref{FigR}. 

\subsection{Relativistic vortex wave packet}

A Lorentz invariant wave function of a scalar packet with OAM can be written as
\begin{eqnarray}
& \displaystyle \psi_{\ell}(p) = \frac{2^{3/2} \pi}{\sigma^{|\ell| + 1}\sqrt{|\ell|!}}\, p_{\perp}^{|\ell|}\,\frac{e^{-m^2/\sigma^2}}{\sqrt{K_{|\ell| + 1} (2m^2/\sigma^2)}}\cr
& \displaystyle \times\exp\left\{\frac{(p - \bar{p})^2}{2\sigma^2} + i\ell\phi_p\right\} \simeq \left(\frac{2\sqrt{\pi}}{\sigma}\right)^{3/2}\,\sqrt{2m}\cr
& \displaystyle \times \frac{p_{\perp}^{|\ell|}}{\sigma^{|\ell|}\sqrt{|\ell|!}}\left (1 - \frac{4(|\ell|+1)^2 - 1}{32}\frac{\sigma^2}{m^2}\right) \cr
& \displaystyle \times \exp\left\{\frac{(p - \bar{p})^2}{2\sigma^2} + i\ell\phi_p\right\},\cr
& \displaystyle \int \frac{d^3p}{(2\pi)^3}\frac{1}{2\varepsilon}\,|\psi_{\ell} (p)|^2 = 1,
\label{OAMrelp}
\end{eqnarray}
and it is no longer real. These states are orthogonal in OAM, 
$$
\int \frac{d^3p}{(2\pi)^3}\frac{1}{2\varepsilon}\,\left[\psi_{\ell^{\prime}} (p)\right]^* \psi_{\ell} (p) = \delta_{\ell,\ell^{\prime}}.
$$

A Fourier transform of this non-paraxial function can also be found exactly,
\begin{eqnarray}
& \displaystyle \psi_{\ell}(x) =  \int \frac{d^3 p}{(2\pi)^3} \frac{1}{2\varepsilon}\, \psi_{\ell}(p) e^{-ipx} = \cr
& \displaystyle = \frac{(i\rho)^{|\ell|}}{\sqrt{2|\ell|!}\, \pi}\,\frac{\sigma^{|\ell| + 1}}{\varsigma^{|\ell|+1}}\,\frac{K_{|\ell|+1} (\varsigma m^2/\sigma^2)}{\sqrt{K_{|\ell|+1}(2m^2/\sigma^2)}}\, e^{i\ell\phi_r},
\label{OAMrelexact}
\end{eqnarray}
with $\varsigma$ from (\ref{psibosonx}), which does not depend on the azimuthal angle $\phi_r$ for a packet moving along the $z$ axis on average. 
That is why for such a packet $\langle \hat{L}_z\rangle = \ell$ where $\hat{L}_z = -i\partial/\partial\phi_p$ or $\hat{L}_z = -i\partial/\partial\phi_r$, depending on the representation.

Regardless of the OAM, the wave function (\ref{OAMrelexact}) still decays exponentially at large distances, $\sqrt{-x_{\mu}^2} \gg \lambda_c$, according to (\ref{psiass}).
Note that within this class of functions, it is the only law allowed by the invariance considerations and it is in sharp contrast with Eq.(15) in Ref.\cite{Birula} 
where the analogous scale is $\bar{\varepsilon}/m$ times smaller than $\lambda_c$, which is impossible within a one-particle approach with a stable vacuum (see, for instance, \cite{BLP}) 
and is not Lorentz invariant. 

An approximate paraxial wave function in the coordinate representation can easily be guessed by comparing Eqs.(\ref{psibosonparaxx}) and (\ref{psiellx}):
\begin{eqnarray}
& \displaystyle \psi_{\ell}^{\text{par}}(x) = \frac{(i\rho)^{|\ell|}}{\sigma^{|\ell|} \sqrt{|\ell|!}}\frac{e^{i\ell\phi_r}}{(1/\sigma^2 + it/\bar{\varepsilon})^{|\ell|}}\, \psi^{\text{par}} (x),\cr
& \displaystyle \int d^3x\,2\bar{\varepsilon}\,|\psi_{\ell}^{\text{par}}(x)|^2 = 1,
\label{OAMrelparaxx}
\end{eqnarray}
where $\psi^{\text{par}} (x)$ is from Eq.(\ref{psibosonparaxx}). This state is also a Lorentz scalar and it represents a relativistic Laguerre-Gaussian beam in the ground state $n=0$ 
with a Gouy phase depending on time rather than on $z$ (see Sec.\ref{LG} for more detail). In the non-relativistic limit we return to Eq.(\ref{psiellx}).

By analogy to (\ref{energymeanex}), one can calculate the vortex packet's mean energy and momentum non-perturbatively,
\begin{eqnarray}
& \displaystyle \langle p^{\mu}_{\ell}\rangle = \{\langle \varepsilon_{\ell}\rangle,\langle {\bm p}_{\ell}\rangle\} = \{\bar{\varepsilon},\bar{\bm p}\}\, \frac{K_{|\ell|+2}\left(2m^2/\sigma^2\right)}{K_{|\ell|+1}\left(2m^2/\sigma^2\right)} \simeq \cr
& \displaystyle \simeq \{\bar{\varepsilon},\bar{\bm p}\}\,\left(1 + \left(\frac{3}{4} + \frac{|\ell|}{2}\right) \frac{\sigma^2}{m^2}\right),
\label{energymeanexvortex}
\end{eqnarray}
 and find that the non-paraxial correction now depends on the OAM and it is also positive. As a result, the invariant mass,
\begin{eqnarray}
& \displaystyle m_{\ell}^2 = \langle p_{\ell}\rangle^2 \simeq m^2\,\left(1 + \left(\frac{3}{2} + |\ell|\right) \frac{\sigma^2}{m^2}\right),
\label{invmassvortex}
\end{eqnarray}
depends on $|\ell|$ too. One can say that a vortex packet is heavier than an ordinary Gaussian beam,
\begin{eqnarray}
& \displaystyle \frac{\delta m_{\ell}}{m_{\text{inv}}} \equiv \frac{m_{\ell} - m_{\ell=0}}{m_{\ell=0}} =\cr
& \displaystyle = \frac{K_{|\ell|+2}\left(2m^2/\sigma^2\right)}{K_{|\ell|+1}\left(2m^2/\sigma^2\right)} \frac{K_{1}\left(2m^2/\sigma^2\right)}{K_{2}\left(2m^2/\sigma^2\right)} - 1 = \cr
& \displaystyle = \frac{|\ell|}{2} \frac{\sigma^2}{m^2} + \mathcal O (\sigma^4/m^4).
\label{deltam}
\end{eqnarray}
For beams with $|\ell| \sim 10^3$ and focused to a spot of $\sigma_{\perp} \gtrsim 0.1$ nm, we have
$$
\frac{\delta m_{\ell}}{m_{\text{inv}}} \simeq \frac{\delta m_{\ell}}{m} \lesssim 10^{-3}.
$$

Thus, the non-paraxial corrections are $|\ell|$ times \textit{enhanced} for highly twisted particles compared to the packets with the non-singular phases 
(say, the Airy beams). To put it differently, it is no longer the Compton wave length $\lambda_c$ that defines a paraxial scale for massive particles with phase vortices but it is
$$
\sqrt{\ell}\,\lambda_c,
$$
which can be more than an order of magnitude larger than $\lambda_c$ for available electrons.

In the paraxial regime, $|\psi_{\ell}^{\text{par}}(x)|^2$ plays a role of the probability density (according to (\ref{OAMrelparaxx})) and its transverse part,
\begin{eqnarray}
& \displaystyle
|\psi_{\ell}^{\text{par}}({\bm \rho}, z = t = 0)|^2 =\text{const}\, \frac{(\rho \sigma)^{2|\ell|}}{|\ell|!}\, e^{-(\rho\sigma)^2},
\label{j0paraxx}
\end{eqnarray} 
is the gamma distribution with the typical doughnut-like profile with a maximum, exactly as in the non-relativistic case. 
It is also because of this Poissonian behavior that the paraxial vortex packets (\ref{OAMrelparaxx}) 
closely resemble coherent states of a quantum oscillator. Unlike the latter, however, these states do not minimize the coordinate-momentum uncertainty relations 
(only the Gaussian packet (\ref{psibosonparaxx}) at $t=0$ does), see also Sec.\ref{LG}. 

For small $\rho \sigma$, the behavior of (\ref{j0paraxx}) is similar to the Bessel beam for which $|\psi_{\ell}(x)|^2 = \text{const}\, J_{|\ell|}^2 (\rho p_{\perp}) \propto (\rho p_{\perp})^{2|\ell|}$ when $\rho p_{\perp} \ll 1$. Thus, the packet width $\sigma$ plays a role of the transverse momentum $p_{\perp}$. 
According to the properties of the gamma distribution, a mathematical expectation for $\sigma^2\rho^2$ is $|\ell|$, 
which gives the $\sqrt{|\ell|}$-scaling of the intensity maximum $\rho_0$, 
\begin{eqnarray}
& \displaystyle \rho_0 \approx \sqrt{|\ell|}/\sigma,
\label{rho0}
\end{eqnarray}
discussed, for instance, in Ref.\cite{Krenn}.

\begin{figure}[t]
	\center
	\includegraphics[width=0.9\linewidth]{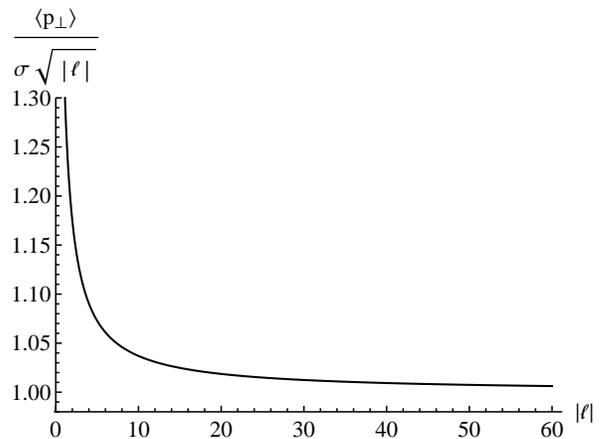}
	\caption{The mean absolute value of the transverse momentum of the vortex packet (\ref{OAMrelp}) in the paraxial regime, $|\ell|\sigma^2/m^2 \ll 1$.
\label{Figell}}
\end{figure}

One can also calculate the packet's mean transverse momentum exactly,
\begin{eqnarray}
& \displaystyle \langle p_{\perp}\rangle = \sigma \frac{\Gamma (|\ell| + 3/2)}{\Gamma (|\ell| + 1)}\,\frac{K_{|\ell| + 3/2}(2m^2/\sigma^2)}{K_{|\ell| + 1}(2m^2/\sigma^2)} \simeq \cr
& \displaystyle \simeq \sigma\,\frac{\Gamma (|\ell| + 3/2)}{\Gamma (|\ell| + 1)}\,\left(1 + \frac{5 + 4|\ell|}{16}\frac{\sigma^2}{m^2}\right ),
\label{pperpmean}
\end{eqnarray}
where the function 
\begin{eqnarray}
& \displaystyle
\langle p_{\perp}\rangle/\sigma \simeq \frac{\Gamma (|\ell| + 3/2)}{\Gamma (|\ell| + 1)} \approx \sqrt{|\ell|}\,\, \text{when}\,\, |\ell| \gg 1,
\label{pperpmean2}
\end{eqnarray}
grows as $\sqrt{|\ell|}$ for high OAM, see Fig.\ref{Figell}. Say, for $|\ell| \sim 10^3$ we have $\langle p_{\perp}\rangle \simeq \sqrt{10^3}\sigma \sim 30\sigma$.
As a result,
\begin{eqnarray}
& \displaystyle \rho_0 \langle p_{\perp}\rangle \approx |\ell|,\, \text{when}\, |\ell| \gg 1,
\label{rho0}
\end{eqnarray}
analogously to the main peak of the Bessel beam. 

The invariant condition of paraxiality can now be rewritten as follows:
\begin{eqnarray}
& \displaystyle |\ell|\frac{\sigma^2}{m^2} \simeq \frac{\langle p_{\perp}\rangle^2}{m^2} = \left(\frac{\bar{\varepsilon}^2}{m^2} - 1\right) \tan^2\theta_0 \ll 1,
\label{ineqnonpar}
\end{eqnarray}
where we have introduced an opening angle $\theta_0 = \arctan \langle p_{\perp}\rangle/\bar{p}$, 
which \textit{grows with} $|\ell|$. One can say that enhancement of the non-paraxial effects for vortex beams owes to their large transverse momentum or to the large opening angle $\theta_0$.
We would like to emphasize that this enhancement cannot be reproduced with the simplified Bessel beams or with the Laguerre-Gaussian packets,
as the former have a definite transverse momentum $\varkappa$, which is independent of $\ell$, while the latter yield just $\langle p_{\perp}\rangle = \sigma \sqrt{|\ell|}$ without the non-paraxial correction.

\section{Relativistic fermion}

One can define a wave function in momentum representation for a wave packet of a fermion as follows:
\begin{eqnarray}
& \displaystyle 
\psi_f (p) = \frac{u(p)}{\sqrt{2\varepsilon}}\, \psi(p),\cr
& \displaystyle \int \frac{d^3p}{(2\pi)^3}\frac{1}{2\varepsilon}\,|\psi_f (p)|^2 = 1 = \text{inv},
\label{WFf}
\end{eqnarray}
where $\psi(p)$ is a normalized bosonic wave function from the previous section (say, (\ref{psiboson}) or (\ref{OAMrelp})) and 
\begin{eqnarray}
& \displaystyle 
u(p) = \left (\sqrt{\varepsilon + m}\,\omega, \sqrt{\varepsilon - m}\,({\bf p}{\bm \sigma})\omega/|{\bf p}|\right )^{T}
\label{up}
\end{eqnarray}
is a bispinor, which obeys $|u(p)|^2 = 2\varepsilon$.
In what follows, we deal with the helicity states for which a 2-component spinor $\omega$ obeys
\begin{eqnarray}
\displaystyle 
({\bf n}{\bm \sigma})\, \omega = 2\lambda\, \omega,\ {\bf n} = \frac{\bar{\bf p}}{|\bar{\bf p}|} \equiv \hat{\bm z},\, \lambda = \pm 1/2,\, \omega^{\dagger}\omega = 1,
\label{lambda}
\end{eqnarray}
where ${\bm \sigma}$ are the Pauli matrices. Note that we project the spin onto the packet's ``mean momentum'' $\bar{\bf p}$ and not onto ${\bf p}$, which is an integration variable.

A wave function in the configuration space, which is an exact solution to the Dirac equation, can be defined as follows (compare this with Eq.(\ref{psibosonx}) for a boson):
\begin{eqnarray}
& \displaystyle 
\psi_f (x) = \int\frac{d^3p}{(2\pi)^3}\frac{1}{\sqrt{2\varepsilon}}\, \psi_f (p) e^{-ipx} = \cr
& \displaystyle = \int\frac{d^3p}{(2\pi)^3}\,\frac{u(p)}{2\varepsilon}\, \psi (p) e^{-ipx}.
\label{WFfx}
\end{eqnarray}
For the Gaussian packets studied in this paper, the integral can be evaluated via the steepest descent method. The resultant paraxial wave function, however, is of little use
and it is more convenient to work in the p-space. 

The 4-current for the fermion is 
\begin{eqnarray}
& \displaystyle 
j = \{|\psi_f (x)|^2, \bar{\psi}_f(x){\bm \gamma}\psi_f(x)\},
\label{je}
\end{eqnarray}
and therefore normalization in the coordinate space is also invariant,
\begin{eqnarray}
& \displaystyle 
\int d^3 x\, j^0 = \int \frac{d^3p}{(2\pi)^3}\frac{1}{2\varepsilon}\,|\psi_f (p)|^2 = 1.
\label{fnorm}
\end{eqnarray}

If $\psi(p)$ in the right-hand side of (\ref{WFfx}) is from (\ref{OAMrelp}), the function $\psi_f (x)$ describes a vortex electron wave packet
with a total angular momentum 
$$
\langle\hat{j_z}\rangle = \ell + \lambda,
$$ 
as can be easily shown by acting by an operator $\hat{j_z} = \hat{L}_z + \hat{s}_z$ on the function $\psi_{f} (x)$ (here $\hat{s}_z = 1/2\,\text{diag}(\sigma_3, \sigma_3)$\cite{BLP}).
Clearly, the mean energy and momentum of this state coincide with the scalar expressions (\ref{energymeanexvortex}).

We now turn to the calculation of an intrinsic magnetic moment 
for an arbitrary wave packet, which is
\begin{widetext}
\begin{eqnarray}
& \displaystyle 
{\bm \mu}_f = \frac{1}{2} \int d^3 x\, {\bf r} \times \bar{\psi}_f(x){\bm \gamma}\psi_f(x) = \cr
& \displaystyle  = \frac{1}{2} \int \frac{d^3p}{(2\pi)^3}\frac{d^3k}{(2\pi)h^3}\,d^3 x\,\frac{\psi^*({\bf p} - {\bf k}/2) \psi({\bf p} + {\bf k}/2)}{2\varepsilon({\bf p} - {\bf k}/2) 2\varepsilon({\bf p} + {\bf k}/2)}\,\, {\bf r}\times \bar{u}({\bf p} - {\bf k}/2){\bm \gamma} u({\bf p} + {\bf k}/2)\cr 
& \displaystyle \times \exp\left\{-it[\varepsilon ({\bf p} + {\bf k}/2) - \varepsilon ({\bf p} - {\bf k}/2)] + i{\bf r} {\bf k}\right\} = \cr
& \displaystyle  = \frac{1}{2} \int \frac{d^3p}{(2\pi)^3}\,d^3k\, \delta({\bf k})\,\, i\frac{\partial}{\partial {\bf k}} \times \bar{u}({\bf p} - {\bf k}/2){\bm \gamma} u({\bf p} + {\bf k}/2)\,\,
\frac{\psi^*({\bf p} - {\bf k}/2) \psi({\bf p} + {\bf k}/2)}{2\varepsilon({\bf p} - {\bf k}/2) 2\varepsilon({\bf p} + {\bf k}/2)},
\label{muf}
\end{eqnarray}
\end{widetext}
where we have taken into account that $\bar{u}({\bf p}){\bm \gamma} u({\bf p}) = 2{\bf p}$ and the time-dependent term has vanished identically. 
Thus ${\bm \mu}_f$ is an integral of motion. Then we apply the following relation for the helicity states (we do not distinguish the upper- and the lower indices here):
\begin{eqnarray}
& \displaystyle 
\bar{u}(p)\,\gamma_j\,\frac{\partial u(p)}{\partial  p_k} - \left (\frac{\partial \bar{u}(p)}{\partial  p_k} \right )\gamma_j\, u(p) = \cr
& \displaystyle = 2i \left (\frac{p_k}{\varepsilon}\, \frac{1}{\varepsilon + m}\, [{\bm \zeta} \times {\bf p}]_j + [{\bm \zeta} \times \hat{{\bf j}}]_k \right ),\ j,k=1,2,3,\cr 
& \displaystyle {\bm \zeta} = 2\lambda {\bf n},
\label{mue1}
\end{eqnarray}
where $\hat{{\bf j}}, |\hat{{\bf j}}| = 1,$ is a unit vector of the $j$-th axis. 

After some calculations, we arrive at the following averages:
\begin{eqnarray}
& \displaystyle 
{\bm \mu}_f = \int \frac{d^3 p}{(2\pi)^3}\,\frac{|\psi(p)|^2}{(2\varepsilon)^3} \left ({\bm \zeta}(\varepsilon + m) + \frac{{\bf p} ({\bf p {\bm \zeta}})}{\varepsilon + m}\right) + \cr
& \displaystyle + \frac{1}{2} \int \frac{d^3 p}{(2\pi)^3}\,\frac{1}{(2\varepsilon)^2}\,\, i{\bm p} \times \Big (\psi(p)\frac{\partial \psi^*(p)}{\partial {\bf p}} - \cr
& \displaystyle - \psi^*(p)\frac{\partial \psi(p)}{\partial {\bf p}}\Big ) = \Bigg \langle \frac{1}{(2\varepsilon)^2} \Big ({\bm \zeta}(\varepsilon + m) + \cr
& \displaystyle  + \frac{{\bf p} ({\bf p {\bm \zeta}})}{\varepsilon + m}\Big)\Bigg\rangle +
\frac{1}{2} \left \langle {\bm u} \times \frac{\partial \varphi(p)}{\partial {\bm p}}\right \rangle \equiv {\bm \mu}_{s} + {\bm \mu}_{b},
\label{muf12}
\end{eqnarray}
where the definition of a mean value $\langle...\rangle$ is from Eq.(\ref{energymean}) and $\varphi(p)$ is a phase of the bosonic wave function,  
\begin{eqnarray}
& \displaystyle 
\psi (p) \equiv |\psi (p)|\,e^{i\varphi(p)}.
\label{varphi}
\end{eqnarray}
The first term in (\ref{muf12}) ${\bm \mu}_{s}$ describes a spin contribution to the magnetic moment, while the second (purely orbital) one ${\bm \mu}_{b}$ represents magnetic moment of a boson 
and it is non-vanishing for packets with phases only. It might seem that both the contributions, that of the phase and that of the spin, do not mix, that is, there is no spin-orbit coupling. 
It is indeed the case when the phase $\varphi(p)$ has no singularities (say, for the Airy beams), but for vortex packets the square of the wave function $|\psi_{\ell}(p)|^2$ still depends on $\ell$ 
and that is why the terms like $\ell{\bm \zeta}$ survive in ${\bm \mu}_{s}$.

Up until now, we did not specify the model of the wave packet and the only assumption we made was (\ref{lambda}). 
Taking then the non-paraxial vortex states (see (\ref{OAMrelp})), one can approximately evaluate the orbital integral in (\ref{muf12}) 
via the steepest descent method, which yields the following result in the laboratory frame of reference:
\begin{eqnarray}
& \displaystyle 
{\bm \mu}_{b} = \frac{1}{2} \left \langle {\bm u} \times \frac{\partial \varphi_{\ell}(p)}{\partial {\bm p}}\right \rangle = \hat{{\bm z}}\, \ell \left\langle \frac{1}{2\varepsilon}\right\rangle \simeq \cr
& \displaystyle \simeq \hat{{\bm z}}\, \ell\, \frac{1}{2\bar{\varepsilon}} \left (1 - \frac{\sigma^2}{2m^2} \left (|\ell| + \frac{1}{2} + \frac{m^2}{\bar{\varepsilon}^2}\right )\right ).
\label{muf2app}
\end{eqnarray} 

A couple of comments on this formula are in order. Unlike the correction to the energy, the one to the magnetic moment is negative.
Neglecting this correction, we return to the result by Bliokh et al., ${\bm \mu}_{b} = \hat{{\bm z}}\, \ell/2\bar{\varepsilon}$ \cite{Bliokh, Review}.
The non-paraxial term is of the order of $|\ell|\sigma^2/m^2$, exactly as with the other observables, even though the integral itself also brings the terms $|\ell|^2\sigma^2/m^2 \gg |\ell|\sigma^2/m^2$.
It is the corresponding $\ell^2$-summand in the expansion of the normalization constant $1/K_{|\ell| + 1}(2m^2/\sigma^2)$ from (\ref{OAMrelp}) that cancels this term.

More interestingly, the factor at the small parameter $\sigma^2/m^2$ turns out to be \textit{not} Lorentz invariant, as it depends on the energy. 
Mathematically, this happens because the function that is to be averaged, $1/2\varepsilon$, does not transform as a component of a tensor 
(cf. Eq.(\ref{energymeanexvortex}) with the energy being a $p^0$).
In the ultrarelativistic regime with $\bar{\varepsilon} \approx \bar{p} \gg m$, the non-invariant term $m/\bar{\varepsilon}$ represents a ratio of the particle's de Broglie wave length, 
$\lambda_{dB} \sim 1/\bar{p}$, to its Compton wave length $\lambda_c$. The standard interpretation of smallness of this ratio is that the particle motion becomes quasi-classical 
in the relativistic regime ($\mathcal O(\hbar^2)$-terms can be neglected). As can be seen, however, there are also the terms at $\sigma^2/m^2$ that are invariant and, 
therefore, do not vanish when $\bar{\varepsilon} \gg m$. In other words, in contrast to the packets with a non-singular phase, the purely quantum corrections influence the vortex packet's motion even in the relativistic case.

Thus, the non-paraxial corrections can be \textit{frame-dependent} for some observables. 
In fact, a pair of dipole moments $({\bf d}, {\bm \mu})$ transforms as a product of an anti-symmetric $4$-tensor and a volume $V$. As the latter is not invariant, 
it is $\varepsilon {\bm \mu}$, not ${\bm \mu}$, that transforms as a component of a tensor. 
Obviously, it is the reason for non-invariance of the correction to the magnetic moment.


The calculations for the spin contribution are more challenging and the result is
\begin{eqnarray}
& \displaystyle 
{\bm \mu}_{s} = \left \langle \frac{1}{(2\varepsilon)^2} \left ({\bm \zeta}(\varepsilon + m) + \frac{{\bf p} ({\bf p {\bm \zeta}})}{\varepsilon + m}\right)\right\rangle \simeq \cr
& \displaystyle \simeq {\bm \zeta}\, \frac{1}{2\bar{\varepsilon}} \Bigg (1 - \frac{\sigma^2}{2m^2} \Big [\frac{1}{2} + \frac{3}{2}\frac{m}{\bar{\varepsilon}} + \frac{1}{2}\frac{m^2}{\bar{\varepsilon}^2} - \frac{3}{2}\frac{m^3}{\bar{\varepsilon}^3} - \cr
& \displaystyle - \frac{m}{\bar{\varepsilon} + m} \left (\frac{3}{2} - 2\frac{m^2}{\bar{\varepsilon}^2} - \frac{3}{2}\frac{m^3}{\bar{\varepsilon}^3} \right ) + \cr
& \displaystyle + |\ell| \left(1 + \frac{m}{\bar{\varepsilon}} - \frac{m}{\bar{\varepsilon} + m}\right ) \Big ]\Bigg )
\label{mufsapp}
\end{eqnarray}
where the two last frame-dependent summands represent a parameter $\Delta$ used for characterizing the spin-orbit coupling in Refs.\cite{Bliokh, Bliokh17, Review},
\begin{eqnarray}
& \displaystyle 
\Delta = \left(1 - \frac{m}{\bar{\varepsilon}}\right) \sin^2\theta_0 \simeq |\ell|\frac{\sigma^2}{m^2} \left(\frac{m}{\bar{\varepsilon}} - \frac{m}{\bar{\varepsilon} + m}\right), 
\label{Delta}
\end{eqnarray}
with the only difference that now it grows with $|\ell|$.
Here, the $|\ell|^2\sigma^2/m^2$ corrections have also been canceled by the corresponding term from the normalization constant and there has appeared the spin-orbit interaction term, 
$$
{\bm \zeta}|\ell|\,\frac{\sigma^2}{m^2} \equiv {\bm \zeta}|\ell|\,\frac{\lambda_c^2}{\sigma_{\perp}^2},
$$
which is also $|\ell|$ times enhanced compared to the Bessel beam (cf. Eq.(20) in \cite{Bliokh}).

As the total magnetic moment of a fermion ${\bm \mu}_f$ represents a sum of (\ref{muf2app}) and (\ref{mufsapp}), 
the spin-orbit coupling will be obscured by the large orbital contribution, which is $\ell$ times stronger.
As a result, although the spin-orbit effects are enhanced for beams with high OAM, the detection of them seems hardly feasible in near future (in accordance with 
the analysis made for Bessel beams in \cite{Boxem}). Note that separation of the magnetic moment into the orbital part and the spin one is unique, as ${\bm r}$ in the left-hand side of (\ref{muf}) is not an operator (see a recent discussion in \cite{Bliokh17}).

Then, similarly to (\ref{muf}), one can also calculate an electric dipole moment,
\begin{eqnarray}
& \displaystyle 
{\bm d}_f = \int d^3 x\, {\bm r}\, j^0 = \left \langle {\bm u} t - \frac{\partial \varphi (p)}{\partial {\bm p}} + \frac{{\bm p}\times {\bm \zeta}}{2\varepsilon (\varepsilon + m)} \right \rangle. 
\label{df}
\end{eqnarray}
where we have used that 
$$
u^{\dagger}(p)\frac{\partial u(p)}{\partial {\bm p}} - \text{c.c.} = \frac{2i}{\varepsilon + m}\, {\bm \zeta}\times{\bm p}.
$$
Due to azimuthal symmetry of the vortex states, two last terms in (\ref{df}) vanish and so ${\bm d}_f = \langle {\bm u}\rangle t = 0$ at $t=0$. 
Note that one can define a mean path of the electron wave packet via its dipole moment as 
$$
\displaystyle \langle {\bm r} \rangle := \frac{{\bm d}_f}{\int d^3 x\, j^0} = \langle {\bm u}\rangle t
$$
where the mean velocity $\langle {\bm u}\rangle = \bar{{\bm u}} (1 + \mathcal O(|\ell|\sigma^2/m^2))$ also acquires the frame-dependent corrections, analogously to (\ref{muf2app}), 
as it is ${\bm u}\,\varepsilon/m$ that represents a spatial component of a $4$-velocity, not ${\bm u}$.


\section{Laguerre-Gaussian beams with $n \ne 0$}\label{LG}

Neglecting the spin, a well normalized packet with OAM and $n+1$ radial maxima ($n\geq 0$) 
is the invariant Laguerre-Gaussian beam with the following paraxial wave function:
\begin{eqnarray}
& \displaystyle 
\psi_{\ell,n}^{\text{par}}(p) = \sqrt{\frac{n!}{(|\ell| + n)!}}\,\left(\frac{2\sqrt{\pi}}{\sigma}\right)^{3/2} \sqrt{2m}\, \left(\frac{p_{\perp}}{\sigma}\right)^{|\ell|}\cr
& \displaystyle \times L_n^{|\ell|}(p_{\perp}^2/\sigma^2)\, \exp\left\{i\ell\phi_p - \frac{p_{\perp}^2}{2\sigma^2} - \frac{m^2}{\bar{\varepsilon}^2}\frac{(p_z - \bar{p})^2}{2\sigma^2}\right\},\cr
& \displaystyle \int\frac{d^3p}{(2\pi)^3}\,\frac{1}{2\varepsilon} |\psi_{\ell,n}^{\text{par}}(p)|^2 = 1,
\label{LGp}
\end{eqnarray}
where $L_n^{|\ell|}$ are generalized Laguerre polynomials. Here we have already neglected the $|\ell|\sigma^2/m^2$ corrections, 
and that is why $1/2\varepsilon = 1/2\bar{\varepsilon}$ within this accuracy. 
For the same reason, these states are monochromatic\footnote{As the non-paraxial corrections $\mathcal O(\sigma^2/m^2)$ also originate from the normalization constant (as (\ref{OAMrelp}) shows)
and we have already neglected them in (\ref{LGp}), the paraxial states cannot be used for calculating the $\sigma^2$-corrections to observables.}, $\langle\varepsilon_{\ell}\rangle = \bar{\varepsilon}$. They are also orthogonal in $n$ and $\ell$,
\begin{eqnarray}
& \displaystyle 
\int\frac{d^3p}{(2\pi)^3}\,\frac{1}{2\varepsilon} \left[\psi_{\ell^{\prime},n^{\prime}}^{\text{par}}(p)\right]^*\psi_{\ell,n}^{\text{par}}(p) = \delta_{n,n^{\prime}}\delta_{\ell,\ell^{\prime}}.
\label{orth}
\end{eqnarray} 

One can also obtain a \textit{non-paraxial} Laguerre-Gaussian beam with $n\ne 0$, analogously to Eq.(\ref{OAMrelp}), which would be an exact solution to the Klein-Gordon equation. 
For such a packet, the radial index $n$ would reveal itself in $n\,\sigma^2/m^2$ corrections. The number of radial maxima discernible in an experiment, however, is usually not large (say, $n \lesssim 10$)
and for the highly twisted beam one can always suppose $|\ell| \gg n$. That is why we shall not study these corrections and restrict ourselves from now on to the paraxial regime.

A Fourier transform of the function (\ref{LGp}) represents a massive generalization of an optical paraxial Laguerre-Gaussian beam,
\begin{widetext}
\begin{eqnarray}
& \displaystyle 
\psi_{\ell,n}^{\text{par}}(x) = \int \frac{d^3p}{(2\pi)^3}\,\frac{1}{2\varepsilon}\, \psi_{\ell,n}^{\text{par}}(p)\,e^{-ipx} = \sqrt{\frac{n!}{(|\ell| + n)!}}\, \frac{i^{2n + \ell}}{\pi^{3/4}\sqrt{2m}}\,\frac{(\rho/\sigma_{\perp}(t))^{|\ell|}}{\sigma_{\perp}^{3/2}(t)}\,
L_{n}^{|\ell|} \left(\rho^2/\sigma_{\perp}^2(t)\right)\cr
& \displaystyle \times \exp\Big\{i\ell\phi_r - i\bar{p}_{\mu} x^{\mu} -i (2n + |\ell| + 3/2)\arctan(t/t_d) -\frac{1}{2\sigma_{\perp}^2(t)}\left(1 - i \frac{t}{t_d}\right)\left (\rho^2 + \bar{\varepsilon}^2(z-\bar{u}t)^2/m^2\right)\Big\},\cr
& \displaystyle 
\int d^3 x\,2\bar{\varepsilon}\, |\psi_{\ell,n}^{\text{par}}(x)|^2 = 1,
\label{LGx}
\end{eqnarray}
\end{widetext}
where
$$
t_d = \frac{\bar{\varepsilon}}{\sigma^2} \sim \frac{\sigma_{\perp}(0)}{\bar{u}_{\perp}},\ \sigma_{\perp}(t) = \sigma^{-1}\sqrt{1 + (t/t_d)^2}
$$
is a diffraction (spreading) time and a beam width, respectively. The latter is Lorentz invariant together with the ratio (recall Eqs.(\ref{psibosonparaxx}),(\ref{rprop}))
$$
t/t_d = \tau\sigma^2/m
$$
Clearly, in the special cases of $n=0$ and $n= \ell =0$ we return to Eq.(\ref{OAMrelparaxx}) and Eq.(\ref{psibosonparaxx}), respectively.

As before, Eq.(\ref{LGx}) is a Lorentz scalar, it does not suffer from the drawbacks discussed in \cite{Birula}, as we do not use the light-cone variables, 
and it has some notable differences from its customary optical counterpart. First, the term 
$$
i (2n + |\ell| + 3/2)\arctan(t/t_d)
$$
represents \textit{an invariant Gouy phase}, which has $3/2$ instead of $1 (\equiv 2/2)$ because the packet is localized in a 3D, not in a 2D, space. 
Then, this phase and the beam width $\sigma_{\perp}(t)$ also depend on the time $t$ and \textit{not} on the distance $z$. 

Indeed, for a paraxial light beam the so-called \textit{time-to-space conversion} takes place, 
$$
z \simeq t.
$$
This is no longer the case, however, for massive particles, especially for beams of the electron microscopes with the energies of hundreds of keV.
While the paraxial states are applicable when $\sigma\ll m$, that is, for a non-vanishing mass,
the ultrarelativistic regime with $\bar{\varepsilon} \approx \bar{p} \gg m$ is correctly described by these formulas\footnote{We stress once again that even in the non-paraxial case, in which a limit $m \rightarrow 0$ can be formally performed, a correct description of the massless case is achieved when $\bar{\varepsilon} \approx \bar{p} \gg m \gtrsim \sigma$ rather than simply putting $m=0$, 
because the one-particle quantum mechanics itself requires that $\sigma$ be a smallest parameter of the theory, not $m$.}
and it is only in this regime that one can substitute $t \rightarrow z$ (due to a saddle point $z = \bar{u}t \simeq t$ in (\ref{LGx})).

To put it differently, it is known that the Klein-Gordon equation can be reduced to the 2-dimensional Schr\"odinger one in the transverse plane 
with the first derivative with respect to time rather than to $z$ (see, for instance, \cite{PRA15, Bagrov, Bagrov_Mono}).
From this, the customary paraxial wave equation with $\partial/\partial z$ is commonly obtained by demanding $p_z \gg p_{\perp}$ \cite{Review, Barnett}.
This inequality, however, is neither invariant nor can it be fulfilled in the non-relativistic domain. 
Indeed, in our notations it is
\begin{eqnarray}
& \displaystyle 
\bar{p} \gg \langle p_{\perp}\rangle \simeq \sigma \sqrt{|\ell|},
\label{ParIneq}
\end{eqnarray}
whereas the invariant condition of paraxiality is $m \gg \sigma \sqrt{|\ell|}$. 
As a result, the Ineq.(\ref{ParIneq}) holds only if $\bar{p} \geq m$, that is, for relativistic particles.

Perhaps the simplest argument why the Gouy phase should depend on $t$ in a Lorentz invariant description is that it is so 
for non-relativistic energies for which the corresponding states (\ref{psiellx}) satisfy the Schr\"odinger equation exactly (of course, Eq.(\ref{LGx}) reduces to Eq.(\ref{psiellx}) in the corresponding limit).
If needed, one can express the Gouy phase in terms of $\langle z\rangle$ by using $t = \langle z\rangle/\bar{u}$, and so $t/t_d = \langle z\rangle\sigma^2/\bar p$.

As we have already noticed, the vortex states are not quasi-classical in a sense that they do not minimize the uncertainty relations, 
and only the Gaussian beams (\ref{psibosonparaxx}) with $n=\ell=0$ at $t=0$ do.
Indeed, calculating the averages, 
\begin{eqnarray}
& \displaystyle 
\langle \rho\rangle = \sigma_{\perp}(t) \frac{\Gamma(n + |\ell| + 3/2)}{\Gamma(n + |\ell| + 1)}, \langle \rho^2\rangle = \sigma_{\perp}^2(t) (n + |\ell| + 1),\cr
& \displaystyle \langle p_{\perp}\rangle = \sigma \frac{\Gamma(n + |\ell| + 3/2)}{\Gamma(n + |\ell| + 1)},\, \langle p_{\perp}^2\rangle = \sigma^2 (n + |\ell| + 1),\cr
& \displaystyle \langle x^2\rangle = \langle y^2\rangle = \frac{1}{2}\,\sigma_{\perp}^2(t) (n + |\ell| + 1),\cr
& \displaystyle \langle p_x^2\rangle = \langle p_y^2\rangle = \frac{1}{2}\,\sigma^2 (n + |\ell| + 1),\cr
& \displaystyle \langle x\rangle = \langle y\rangle = \langle p_x\rangle = \langle p_y\rangle = 0,
\label{rmean}
\end{eqnarray}
we arrive at the following results ($\Delta x = \sqrt{\langle x^2\rangle - \langle x\rangle^2}$):
\begin{eqnarray}
& \displaystyle 
\Delta \rho \Delta p_{\perp} = \sqrt{1 + (t/t_d)^2} \Big(n + |\ell| + 1 - \cr
& \displaystyle - \left(\frac{\Gamma(n + |\ell| + 3/2)}{\Gamma(n + |\ell| + 1)}\right )^2\Big) \approx \cr
& \displaystyle \approx \sqrt{1 + (t/t_d)^2} \left(\frac{1}{4} - \frac{1}{32 |\ell|}\right),\cr
& \displaystyle \frac{\Delta \rho}{\langle\rho\rangle} = \frac{\Delta p_{\perp}}{\langle p_{\perp}\rangle}\approx \frac{1}{2\sqrt{|\ell|}}\ \text{when}\, |\ell| \gg n,\cr
& \displaystyle \Delta x \Delta p_x = \Delta y \Delta p_y = \frac{1}{2}\,\sqrt{1 + (t/t_d)^2} \left(n + |\ell| + 1\right) \approx \cr
& \displaystyle \approx \frac{|\ell|}{2} \sqrt{1 + (t/t_d)^2}
\label{unc}
\end{eqnarray}
when $|\ell| \gg n$. Clearly, $\Delta p_{x,y} \geq \sigma/\sqrt{2}\,;\, \Delta x, \Delta y \geq \sigma^{-1}/\sqrt{2}$, that is, there is no squeezing for any $n, |\ell|$.

More importantly, the beam's mean radius $\langle \rho\rangle$, unlike the beam width $\sigma_{\perp}(t)$, depends on the OAM:
\begin{eqnarray}
& \displaystyle 
\langle \rho\rangle = \sigma_{\perp}(t) \frac{\Gamma(n + |\ell| + 3/2)}{\Gamma(n + |\ell| + 1)} \approx \sigma_{\perp}(t) \sqrt{|\ell|},
\label{beamwidth}
\end{eqnarray}
when $|\ell| \gg n$. Thus, the highly twisted beam is $\sqrt{|\ell|} \gg 1$ times wider than that with $|\ell| \sim 1$, as also noted in Ref.\cite{Krenn}.
To be precise, \textit{one should call} $\langle \rho\rangle = \langle \rho\rangle(t)$, not just $\sigma_{\perp}(t)$, \textit{a width of the vortex beam}. 
When producing vortex particles with the help of the computer generated holograms or the phase plates, $\sigma_{\perp}$ represents the width of the OAM-less beam incident onto a mask or a plate (in-state),
while $\langle \rho\rangle$ defines the width of the diffracted beam with OAM (out-state).

Now let us return to the non-paraxial corrections, which are of the order of 
\begin{eqnarray}
& \displaystyle
|\ell|\frac{\sigma^2}{m^2} \approx \ell^2 \frac{\lambda_c^2}{\langle\rho\rangle^2},
\label{nonparcorr}
\end{eqnarray}
where the beam width $\langle\rho\rangle$ is taken at $t=0$. Although this width grows with the OAM as $\sqrt{|\ell|}$,
the numerator \textit{grows faster}. As a result, for a vortex beam with $|\ell| \gtrsim 10^3$ focused to a spot of 
$\langle\rho\rangle \sim 1 - 10$ nm the conservative estimate of this correction is
$$
\ell^2 \frac{\lambda_c^2}{\langle\rho\rangle^2}\sim 10^{-4} - 10^{-3}.
$$
Note that the mean absolute value of the transverse momentum also grows as $\sqrt{|\ell|}$ and therefore at $t=0$ (cf. Eq.(\ref{rho0}))
\begin{eqnarray}
& \displaystyle 
\langle p_{\perp}\rangle\langle \rho\rangle \approx |\ell|,\ |\ell| \gg n.
\label{transunc}
\end{eqnarray}

\begin{figure}[t]
	\center
	\includegraphics[width=0.95\linewidth]{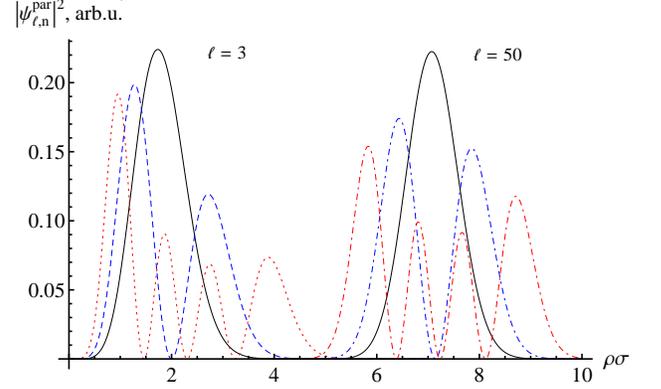}
	\caption{Radial distribution of the Laguerre-Gaussian beam according to Eq.(\ref{LGpsi2}) at $t=0$ for $\ell = 3$ (left), $\ell = 50$ (right). Black: $n=0$ (Poissonian behavior), blue (dashed- and dash-dotted lines): $n=1$, red (dashed- and dash-dotted lines): $n=3$. For $n\ne 0$, the beam as a whole is wider than in the ground state, that is, super-Poissonian. However each line in a set of $n+1$ maxima is slightly narrower than that for $n=0$.
\label{Fignl}}
\end{figure}

When $n\ne 0$, the probability density is no longer Poissonian,
\begin{eqnarray}
& \displaystyle 
|\psi_{\ell,n}^{\text{par}}(x)|^2 \propto \frac{n!}{(|\ell| + n)!}\,\left(\frac{\rho}{\sigma_{\perp}(t)}\right)^{2|\ell|}\cr
& \displaystyle \times \left[L_{n}^{|\ell|} \left(\rho^2/\sigma_{\perp}^2(t)\right)\right]^2\, 
\exp\Big\{- \frac{\rho^2}{\sigma_{\perp}^2(t)}\Big\}.
\label{LGpsi2}
\end{eqnarray}
As can be seen from (\ref{rmean}) and in Fig.\ref{Fignl}, the radial width $\Delta \rho\, \sigma$ of the whole beam is slightly larger for $n\ne 0$ 
than for the fundamental mode $n=0$, which reveals \textit{a super-Poissonian} distribution. On the other hand, each individual line in a set of $n+1$ maxima becomes slightly narrower than that of the ground state (that is, sub-Poissonian).

When the number of radial maxima is very large, $n \rightarrow \infty$, and the packet is wide, $\sigma_{\perp} \rightarrow \infty$, the Laguerre-Gaussian beam turns into the non-diffracting Bessel one (see Eq.8.978 in \cite{Gr}),
\begin{eqnarray}
& \displaystyle 
\psi_{\ell,n}^{\text{par}}(x) \rightarrow \text{const}\, J_{\ell}(2\sigma \rho) \exp\left\{i\ell\phi_r - i\bar{p}_{\mu} x^{\mu}\right\},
\label{Bess}
\end{eqnarray}
where $2\sigma$ plays a role of the transverse momentum $\varkappa$, which is independent of $\ell$ now, and the constant formally diverges, 
as the condition $n \rightarrow \infty$ breaks the paraxial approximation, $n\, \sigma^2/m^2 \ll 1$,
and so the Bessel beam is to be normalized in a cylinder of a large but finite volume.
It is also because of this condition, $n \gg |\ell|$, that the Bessel beam is unsuitable for precise quantitative estimates of the non-paraxial phenomena with large $\ell$.

Finally, Eqs.(\ref{LGp}),(\ref{LGx}) can be generalized for a beam with two different momentum uncertainties, $\sigma_{p,\perp} \ll m$ and $\sigma_{p,z} \ll m$,
which can be the case in experiments. Let 
\begin{eqnarray}
& \displaystyle 
\sigma_{\perp}(t) = \sigma_{p,\perp}^{-1}\sqrt{1 + \left(t/t_{d,\perp}\right)^2},\ t_{d,\perp} = \bar{\varepsilon}/\sigma_{p,\perp}^2,\cr
& \displaystyle 
\sigma_z(t) = \sigma_{p,z}^{-1}\sqrt{1 + \left(t/t_{d,z}\right)^2},\ t_{d,z} = \bar{\varepsilon}/\sigma_{p,z}^2,
\label{widths}
\end{eqnarray}
be the width of the (OAM-less) beam and its length, respectively (both are invariant). The corresponding normalized wave functions are
\begin{widetext}
\begin{eqnarray}
& \displaystyle 
\psi_{\ell,n}^{\text{par}}(p) = \sqrt{\frac{n!}{(|\ell| + n)!}}\,\frac{2\sqrt{\pi}}{\sigma_{p,\perp}}\sqrt{\frac{2\sqrt{\pi}}{\sigma_{p,z}}} \sqrt{2m}\, \left(\frac{p_{\perp}}{\sigma_{p,\perp}}\right)^{|\ell|} 
L_n^{|\ell|}(p_{\perp}^2/\sigma_{p,\perp}^2)\, \exp\left\{i\ell\phi_p - \frac{p_{\perp}^2}{2\sigma_{p,\perp}^2} - \frac{m^2}{\bar{\varepsilon}^2}\frac{(p_z - \bar{p})^2}{2\sigma_{p,z}^2}\right\},\cr
& \displaystyle 
\psi_{\ell,n}^{\text{par}}(x) = \sqrt{\frac{n!}{(|\ell| + n)!}}\,\frac{i^{2n + \ell}}{\pi^{3/4}\sqrt{2m}}\,\frac{(\rho/\sigma_{\perp}(t))^{|\ell|}}{\sigma_{\perp}(t)\sqrt{\sigma_z(t)}}\,
L_{n}^{|\ell|} \left(\rho^2/\sigma_{\perp}^2(t)\right) \exp\Big\{i\ell\phi_r - i\bar{p}_{\mu} x^{\mu} -i (2n + |\ell| + 1)\arctan(t/t_{d,\perp}) - \cr
& \displaystyle - \frac{i}{2} \arctan(t/t_{d,z}) -\frac{1}{2\sigma_{\perp}^2(t)}\left(1 - i \frac{t}{t_{d,\perp}}\right)\rho^2 - \frac{1}{2\sigma_z^2(t)}\left(1 - i \frac{t}{t_{d,z}}\right)\frac{\bar{\varepsilon}^2}{m^2}(z-\bar{u}t)^2\Big\},
\label{LGpperpz}
\end{eqnarray}
\end{widetext}
and when $\sigma_{p,\perp} = \sigma_{p,z} \equiv \sigma$ we return to (\ref{LGp}),(\ref{LGx}). As expected, these states have two diffraction times, $t_{d,\perp}$ and $t_{d,z}$, 
and therefore two invariant Gouy phases. If needed, non-paraxial generalizations of these expressions can also be readily found.

\section{Summary}

Many properties of the relativistic vortex beams, especially the spin-orbit interaction as well as the non-paraxial phenomena in scattering and radiation, 
can be quantitatively treated only when using the spatially localized wave packets described in a Lorentz invariant way and applicable beyond the paraxial approximation.
We have proposed such massive packets and showed that their paraxial counterparts, the invariant Laguerre-Gaussian beams, have notable differences from the corresponding states 
of the twisted photons, especially in the non-relativistic case. As it also turns out, the non-paraxial corrections to observables grow linearly with the OAM for such packets, 
\textit{regardless} of the $\sqrt{|\ell|}$-broadening of them and in contrast to the Bessel beams or to those with the non-singular phases. For vortex electrons with $\sigma_{\perp} \lesssim 1$ nm and $|\ell| > 10^3$ these corrections can reveal themselves in experiments with freely propagating energetic particles, especially in collisions. 

For instance, one of the non-paraxial effects is increase of the invariant mass of the free electron packet by $0.01\%-0.1\%$ due to the high OAM. 
The electron mass can also be retrieved in a tabletop Compton scattering experiment -- see, for instance, \cite{Compton}.
For this effective mass shift to be detected in scattering of optical photons by the moderately relativistic vortex electrons, 
one needs to perform such measurements with a resolution better than $0.1\%$, which seems challenging but feasible with modern technology.
It is also this shift that reveals itself in the corresponding correction $d\sigma^{(1)}$ to the plane-wave cross section $d\sigma_{\text{pw}}$\cite{JHEP}, 
$$
d\sigma^{(1)}/d\sigma_{\text{pw}}\sim |\ell|\sigma^2/m^2 \gtrsim \alpha_{em}^2 = 1/137^2,
$$
Thus, the highly twisted electrons, protons, neutrons, and other massive particles can become a useful tool
for particle physics beyond the conventional plane-wave regime. In order to be applied in hadronic physics, 
the particle's mean transverse momentum should reach the characteristic hadronic scale of at least $\langle p_{\perp}\rangle \sim 10$ MeV.
For the OAM of $\ell \sim 10^3$, one would need to focus the electron beam to a spot of $\langle \rho\rangle \approx |\ell|/\langle p_{\perp}\rangle \sim 0.01$ nm, 
which seems hardly feasible now, whereas for $ \ell \sim 10^5$ the needed requirement is already $\langle \rho \rangle \sim 1$ nm.


The non-paraxial corrections described in this paper, $\sim|\ell|\lambda_c^2/\sigma_{\perp}^2$, can be called kinematic or geometric, 
as there are also \textit{dynamic} effects, which are attenuated as $\lambda_c/\sigma_{\perp}$ but arise only in scattering beyond the Born approximation 
and only if the azimuthal symmetry of the incoming states is broken -- say, when $\langle{\bm p}_{\perp}\rangle \ne 0$ \cite{JHEP}. 
Clearly, such large corrections do not appear in description of the wave packets themselves, as can be easily shown by considering 
a packet with a non-vanishing (but small) mean transverse momentum, $\langle{\bm p}_{\perp}\rangle^2\lesssim\sigma^2 \ll m^2$.

\

I am grateful to V.~Bagrov, I.~Ginzburg, V.~Grillo, I.~Ivanov, D.~Naumov, V.~Serbo, A.~Zhevlakov, and, especially, to P.~Kazinski for many useful discussions and criticism. 
This work is supported by the Russian Science Foundation (project No.\,17-72-20013).

\end{document}